\documentclass{rspublic}

\usepackage[dvips]{graphicx}
\usepackage{amsmath}
\usepackage{amssymb}
\usepackage{latexsym}

\begin{document}

\title[Curvature in conformal mappings of 2D lattices and foam structure]{Curvature in conformal mappings of 2D lattices and foam structure}
\date{\today}

\author[A Mughal \& D Weaire]{Adil Mughal (1,2) and Denis Weaire (3,4)}
\affiliation{(1) ISI Foundation, Viale S. Severo 65, 10133 Torino, Italy (2)Institute of Mathematics and Physics, Aberystwyth University, SY23 3BZ, UK. (3) Universit\'{e} Paris-Est, Laboratoire de Physique des Mat\'{e}riaux Divis\'{e}s et des Interface, UMR CNRS 8108 5 Bd Descartes, 77454 Marne-la-Vall\'{e}e cedex 2, France (4) Department of Physics, Trinity
  College, Dublin 2, Republic of Ireland}

\label{firstpage}
\maketitle

\maketitle
\begin{abstract}{Conformal Crystals, Foams, Curvature}
  The elegant properties of conformal mappings, when applied to two
  dimensional (2D) lattices, find interesting applications in 2D foams
  and other cellular or close packed structures. In particular the 2D
  honeycomb (whose dual is the triangular lattice) may be transformed
  into various conformal patterns, which compare approximately to
  experimentally realisable 2D foams. We review and extend the
  mathematical analysis of such transformations, with several
  illustrative examples. New results are adduced for the
  local curvature generated by the transformation.
\end{abstract}

\section{Introduction}

The relationship $w=f(z)$, where $f$ is any analytical function, can
be viewed as a mapping which sets up a correspondence between the
points of the $z$ and $w$ planes. Such mappings are known as conformal
mappings. The geometrical operations of inversion, reflection,
translation and magnification are all examples of conformal
transformations in Euclidean space.  Conformal mappings have a number
of interesting properties, the most important being isogonality: any
two curves that intersect are transformed into curves that intersect
{\it at the same angle}. 

Consider a discrete 2D set of points in the z-plane generated by two
primitive vectors: the resulting structure in the image plane, due to
a conformal mapping, is known as a {\it conformal lattice}. It is
therefore a purely geometrical object. A {\it strictly conformal
  crystal} is a physical system consisting of particles located on the
sites of a conformal lattice. A {\it conformal crystal} is a physical
system, in which the arrangement of particles approximates a conformal
lattice, see Rothen and Piera\'{n}ski (1996).

There are numerous examples of conformal crystals occurring both in
nature and in the laboratory; see for example Rothen et al (1993,
1996), despite this the geometric properties of conformal crystals are
at present poorly understood. In the present paper we examine some of
the factors which determine the local curvature in these conformal
patterns. We shall begin our analysis by considering the equation for
the complex curvature, which was derived by by Needham (1997) and also
by Mancini and Oguey (2005 a,b). For a given line in the z-plane, we
shall see that whereas the first derivative of the transforming
function relates the direction of a line to its transformed
counterpart, the induced curvature involves the second
derivative. Here we shall use the equation for the induced curvature
to compute the mean and mean square curvature of the conformal lattice

One of the most easily recognised examples of a naturally occurring
conformal crystal is the phyllotactic design of a sunflower, see
Rothen et. al. (1993). Another example is the so called ``gravity's
rainbow'' structure, which is the name given to the striking
arrangement of arches formed by a cluster of magnetised steel balls in
an external force field, see Rothen et. al. (1993). More recently,
conformal lattices have been shown to have a connection with
disclinations in 2D crystalline structures, see Mughal and Moore
(2007) and Riviera et. al. (2005).

Conformal crystals can be physically realised by sandwiching an
ordered, quasi 2D, foam in a Hele-Shaw cell with non-parallel plates,
see for example Drenckhan et. al. (2004). Yet another method involves
the use of ferrofluid foams in magnetic fields, see Elias
et. al.(1999). The advantage of these foam-based methods is that a
variety of conformal crystals can be realised by tuning the geometry
of the experiment. However, the use of foams to approximate conformal
lattices involves two complications: firstly as was shown by Mancini
and Oguey (2005 a, b) the curvature of the soap films {\it
  perpendicular} to the glass plates has to be taken into account,
secondly the total curvature of a soap film must always be
constant. The limitations that these conditions impose on realising a
given conformal crystal, using foams, will be discussed in detail
below.

The paper is organised as follows. In section 2 we introduce the
complex curvature and give some properties of conformal
transformations. The relationship between ordered 2D soap froths and
conformal transformations is detailed in section 3. In section 4 we
calculate the mean curvature and the mean square curvature of the
conformal lattice when the original lattice in the z-plane is free of
curvature. We illustrate these results with
some examples which include the case of complex inversion. In section
5 we generalise our results and include the case where the original
lattice in the z-plane has a curvature. In section 6 we compute the
higher order terms in the expression for the complex curvature.

\section{Some Properties of Conformal Transformations}

\subsection{Scaling of Areas}

Although a conformal mapping $w=f(z)$ preserves angles (the isogonal
property) it does not preserve areas. If
$ds_{z}=(dx^2+dy^2)^{\frac{1}{2}}$ is a small element of line in the
(x,y) plane, upon being mapped to the w-plane it will be magnified
and have a length given by,
\begin{equation}
ds_{w}
=
\left|
\frac{dw}{dz}
\right|
ds_{z}
=
|f_{1}(z)|ds_{z},
\label{eq:lscale}
\end{equation}
where $f_{n}(z)$ is the nth derivative of the function $f(z)$. Hence a
small element of area in the z plane, denoted by $dA_z$, will upon
being mapped to the w-plane have an area
\begin{equation}
dA_{w}
=
\left|
\frac{dw}{dz}
\right|^2
dA_z
=
|f_{1}(z)|^2
dA_{z}.
\label{eq:ascale}
\end{equation}

\subsection{Complex Curvature}

\begin{figure}
\begin{center}
\includegraphics[width=1.0\columnwidth]{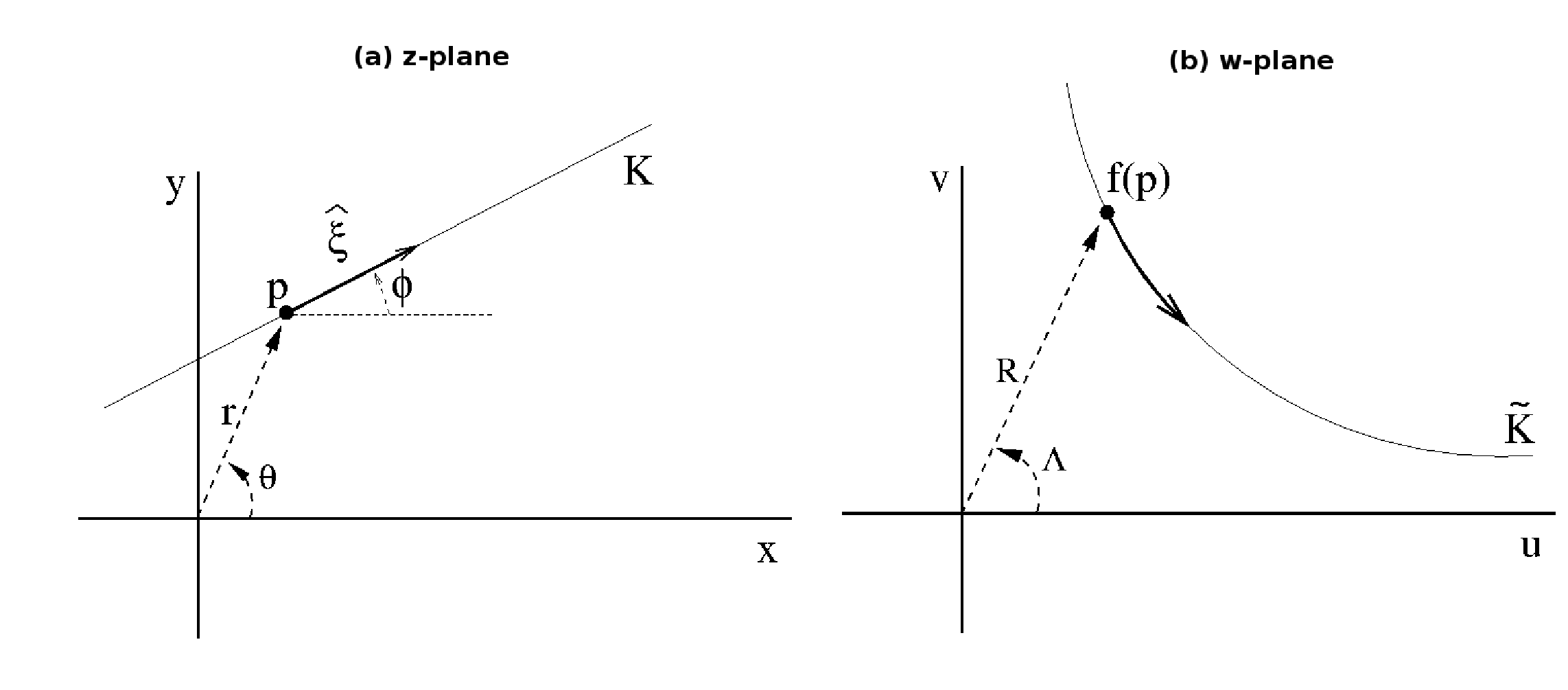}
\caption{{\bf (a)} For simplicity we choose to consider a straight
    line K, in the z-plane, which has no curvature (i.e.  $\kappa=0$).
    As shown the tangent vector $\widehat{\xi (\phi)}$ at the point p
    makes an angle of $\phi$ with the x axis. {\bf (b)} In the
    image plane the effect of the analytical mapping is to yield a
    curve $\widetilde K$, the curvature of which is given by
    $\widetilde{\kappa}$. After Needham (1997)}
\label{complex_curvature}
\end{center}
\end{figure}

Consider a curve K in the z-plane: if we apply an analytical mapping f
to this curve then it will transform into another curve in the image
plane, which we denote by $\widetilde K$. Let us now choose some arbitrary point on K which we denote by
$p=x+iy=re^{i\theta}$. The unit tangent vector to the curve at point p
is given by $\widehat{\xi (\phi)}=e^{i\phi}$, where $\phi$ is the
angle the tangent vector makes with the x axis; for an illustration
see Fig. (\ref{complex_curvature}a). Upon applying an analytical
transformation, point p is mapped to a new point in the image plane
which has coordinates $f(p)=u+iv=Re^{i\Lambda}$. It has been shown,
see Needham (1997), that if the instantaneous curvature of K at the point
p is given by $\kappa$, then the instantaneous curvature at f(p) is
given by,
\begin{equation}
\widetilde{\kappa}
=
\frac{1}{|f_{1}(p)|}
\text{Im}
\left[
\frac{f_{2}(p)}{f_{1}(p)}
\widehat{\xi (\phi)}
\right]
+
\frac{\kappa}{|f_{1}(p)|}
,
\label{eq:curvature}
\end{equation}
where Im is the imaginary component, see Fig.(\ref{complex_curvature}b). The first term in Eq. ($\!\!$~\ref{eq:curvature}) is the
curvature in the image plane if K is a straight line. If however K has
a curvature $\kappa$, then in the image plane this additional
curvature is scaled by a factor of $1/|f_{1}(p)|$; again see Needham (1997)
for a beautiful derivation of these and other results.

The utility of Eq.  ($\!\!$~\ref{eq:curvature}) can be demonstrated by
a short example. Consider the mapping,
\begin{equation}
w=f(z)=z^{\alpha},
\label{eq:example_transformation}
\end{equation}
where
\begin{equation}
f_{1}(z)=\alpha z^{\alpha-1} 
\;\;\;\text{and}\;\;\;
f_{2}(z)=\alpha (\alpha-1) z^{\alpha-2}, 
\label{eq:derivatives}
\end{equation}
and let us represent the z-plane and the image plane in terms of
complex polar coordinates so that,
\[
z=re^{i\theta}
\;\;\;\text{and}\;\;\;
w=Re^{i\Lambda}.
\]
Upon substituting Eq. ($\!\!$~\ref{eq:derivatives}) into Eq.
($\!\!$~\ref{eq:curvature}), we find that the the effect of mapping a
straight line in the z-plane (i.e. $\kappa=0$) is to yield a curve in
the w-plane with curvature,
\[
\widetilde{\kappa}
=
\frac{1}{\alpha r^{\alpha-1}}
\text{Im}
\left[
\frac{\alpha-1}{z}
\widehat{\xi (\phi)}
\right].
\]
Writing out $\widehat{\xi (\phi)}$ explicitly and simplifying, this becomes,
\begin{equation}
\widetilde{\kappa}
=
\frac{1}{\alpha r^{\alpha-1}}
\text{Im}
\left[
\frac{\alpha-1}{re^{i\theta}}
\left( e^{i\phi}  \right)
\right]
=
\frac{\alpha -1}{\alpha}
\frac{1}{r^{\alpha-1}}
\text{Im}
\left[
\frac{e^{i(\phi-\theta)}}{r}
\right]
=
\frac{\alpha -1}{\alpha}
\frac{1}{r^{\alpha}}
\sin(\phi-\theta).
\nonumber
\end{equation}
Note, the curvature is expressed in terms of the coordinates of the
z-plane. To get the curvature in the w-plane we must use the
relationship $R=r^{\alpha}$, which gives
\[
\widetilde{\kappa}
=
\frac{\alpha-1}{\alpha}
\frac{1}{R}
\sin(\phi-\theta).
\] 
Thus the maximum curvature occurs when the tangent vector is
perpendicular to the vector connecting the origin to the point p and
is given by,
\begin{equation}
|\widetilde{\kappa}_{max}|
=
\frac{\alpha-1}{\alpha}
\frac{1}{R}.
\label{eq:maxK}
\end{equation}
This means that lines, in the z-plane, drawn perpendicular to the
vector connecting the origin with point p will acquire the greatest
curvature, while lines drawn parallel to it will not suffer any
curvature.

\section{Properties of 2D Soap Foams and Their
  Relationship to Conformal Transformations}

\begin{figure}[t]
\begin{center}
\includegraphics[width=0.6\columnwidth]{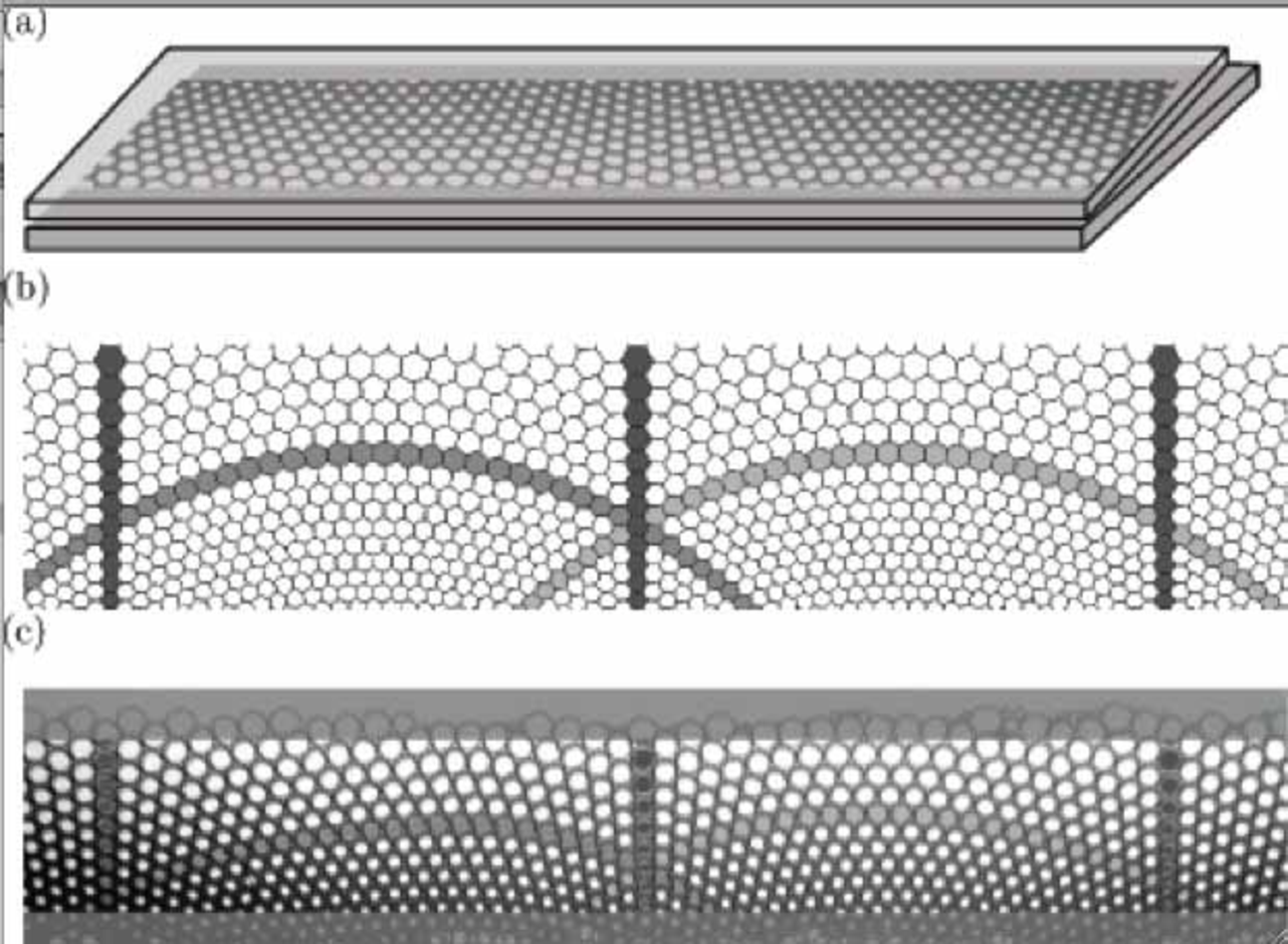}
\caption{From Drenckhan et. al. (2004): a realisation of the logarithmic
  map (also known as the gravity's rainbow structure). {\bf (a)}
  Experimental setup. Monodisperse bubbles are trapped between two
  glass plates, the lower plate is horizontal while the upper plate is
  angled. This arrangement induces specific variation of the bubble
  area with position in the foam. {\bf (b)} Numerical mapping of the perfect
  honeycomb using the transformation $w=(i\alpha)^{-1}\log (i\alpha
  z)$. {\bf (c)} Experimentally obtained pattern. The shading shows how
  lines that were initially straight are transformed.}
\label{stopgap}
\end{center}
\end{figure}

A dry 2D foam consists of 2D bubbles separated by lines which meet at
vertices. Such a foam can be represented by a network consisting of 2D
cells separated by 1D edges. Equilibrium conditions impose strict
restrictions on the topology and geometry of such a foam network (see
Weaire and Hutzler (2001)): Plateau's law stipulates that the edges
can intersect only three at a time and must do so at an angle of $2\pi
/3$; the edges themselves have a constant curvature (circular arcs)
and the curvature is related to the corresponding pressure difference
between the adjacent bubbles by the Laplace-Young relation. It follows
that the sum of the curvatures of the three edges at a given vertex
vanishes, or equivalently that the mean curvature of the adjoining
edges vanishes.

All of the above conditions are automatically satisfied by complex
inversion, $f(z)=1/z$, or more generally a bilinear conformal
transformation which has the form
\[
f(z)=\frac{az+b}{cz+d}.
\]
It can be decomposed into four sequential transformations:
translation, inversion, expansion and rotation, and a final
translation, see Needham (1997). Of these only inversion is
non-trivial. Conformality ensures that the mean curvature at each
vertex vanishes, while only complex inversion (or a bilinear
transformation) has the special property that it will map a circular
arc into another circular arc.  Given a dry 2D foam structure at
equilibrium, inversion will therefore produce a new equilibrium
structure; this was discussed by Weaire (1999) who used it to provide
a neat proof of the Decoration Theorem.

A quasi 2D foam can be realised by sandwiching a single layer of
bubbles between a pair of narrowly separated glass plates(Hele-Shaw
cell).  In the ideal case, if all the bubbles trapped in the Hele-Shaw
cell have the same volume (monodisperse), then an ordered quasi 2D
foam can be realised. Upon viewing the Hele-Shaw cell from above
(i.e. from the direction perpendicular to the plates) the bubbles are
observed to form a honeycomb structure.

Let us now assume that the bottom plate of the Hele-Shaw cell is flat
while the upper plate is slightly angled or curved, if the bubbles are
monodisperse, then this imposes a specific variation in the area of
the bubbles. The variation in the bubble area can be made to closely
match the variation required by a given conformal transformation - as
stipulated by Eq. ($\!\!$~\ref{eq:ascale}). Note that although the
area of the bubbles (as observed from the direction perpendicular to
the bottom plate) may change their volume remains constant, thus the
height of the upper surface is related to the analytical function
$f(z)$ by $h(w)=1/dA_{w}$. This fact has been used to transform the
(straight-edged) honeycomb structure, in a variety of ways, to
generate approximations of conformal lattices (i.e. conformal
crystals) Drenckhan (2004), see for example Fig. (\ref{stopgap}).

In the case of a quasi 2D foam it is important to remember that the
boundaries between bubbles are in fact 2D soap films and not 1D edges,
as they are often approximated. Thus, as noted by Mancini and Oguey
(2005 a,b), the mean curvature of the soap film H is the average of the
transverse curvature $\kappa_t$ (i.e. the curvature of soap film in
the direction perpendicular to the glass plates) and longitudinal
curvature $\kappa$ (i.e. the curvature parallel to the glass
plates),
\[
H=\frac{1}{2}(\kappa+\kappa_t).
\]
Mancini and Oguey considered two cases. The first is the trivial case
when the internal pressure is the same for all bubbles in the
conformal crystal and thus $\kappa_t = -\kappa$ and $H=0$. From an
experemental perspective this case is somewhat artifical. Of more
direct relevance to experimental situations is the second case where
the volume of all the bubbles is the same. In this case it is found
that $\kappa_t = -2\kappa$ and therefore that the mean curvature does
not vanish and is given by $H=-\frac{1}{2}\kappa$. 

Since in the case of constant volume bubbles, in general the
longitudinal curvature $\kappa$ - and therefore the mean curvature H -
is not constant for a given arc it is important to know by how much
the total curvature deviates from being a constant (as required by the
conditions of equilibrium). This discrepancy sets a limit on the
applicability of conformal transformations in the context of foams. In
section VI we calculate the magnitude of this discrepancy to lowest
order.

\section{Lattice Curvature and Conformal Transformations}

In this section we shall examine the curvature that a lattice, in the
z-plane, acquires upon being mapped to the w-plane. Although the
triangular lattice is of primary importance, the results derived are
general enough to include other structures. This includes the square
lattice and the honeycomb structure. To keep things simple we only
consider the case where the original edges in the z-plane, connected
to a given vertex, are free of curvature. This condition will be
relaxed in the next section.

\begin{figure}
\begin{center}
\includegraphics[width=0.4\columnwidth]{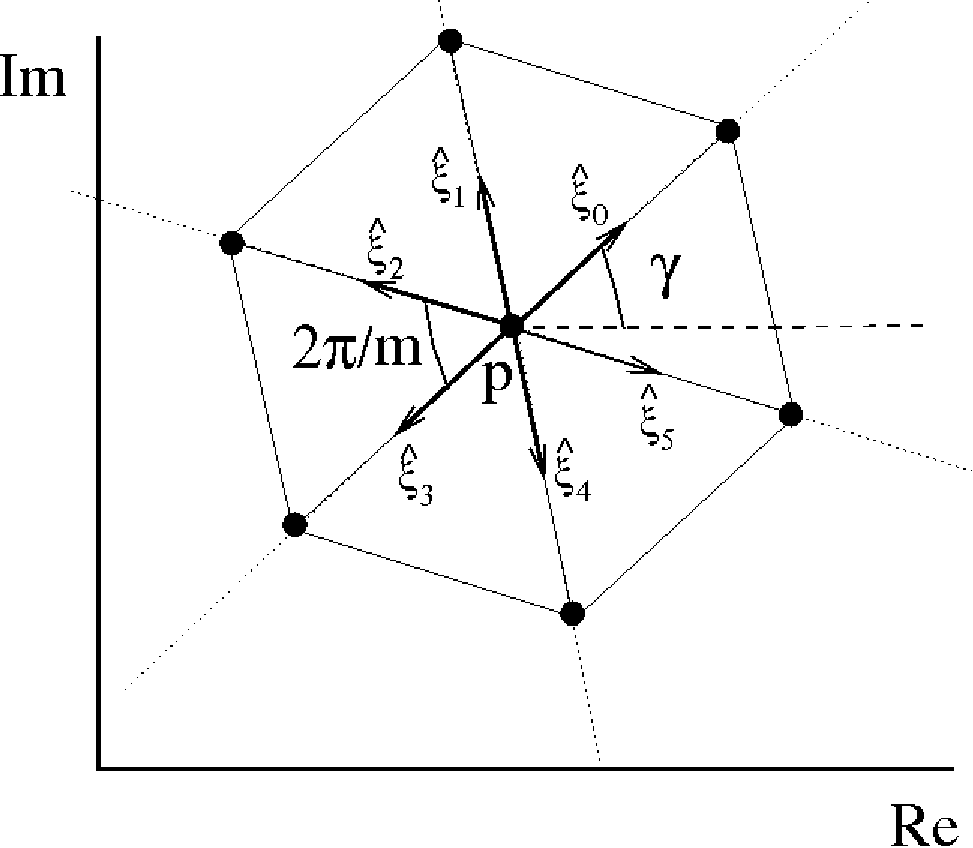}
\caption{Consider a vertex located at point p in the z-plane,
  connected to which there are m straight edges. From the point p we
  can draw tangent vectors along each edge which is denoted by
  $\widehat{\xi_0}$,$\widehat{\xi_1}$...$\widehat{\xi_{m-1}}$. The
  angular separation between the tangent vectors is given by $2\pi/m$.
  We define an angle $\gamma$ which is the angle between the x-axis
  and the first vector $\widehat{\xi_0}$. Thus the angle that the nth
  vector makes with the x axis is $2\pi n/m + \gamma$. Clearly the
  edges are invariant under a rotation of $2\pi /m$ and so we
  restrict $\gamma$ to the range $0 \leq \gamma < 2\pi/m$.}
\label{initial}
\end{center}
\end{figure}

Consider a vertex in the z-plane located at p=x+iy. Connected to this
vertex $m \geq 3$ straight edges labbled $\xi_n$, with $n=0,1,2...m$,
see Fig. (\ref{initial}). The angular separation between the tangent
vectors of successive edges being equal to $2\pi/m$. In the case of
$m=6$ or $m=4$ it is possible to define two edges, which are connected
to the vertex, as primitive vectors and thus tessellate the z-plane
forming a triangular or square Bravais lattice, respectively. We
cannot, however, generate the honeycomb structure in the same way
(using a vertex with $n=3$) since the honeycomb structure is not a
Bravais lattice. Nevertheless we can tessellate the z-plane with a
honeycomb structure, in a unique way, by making a Voronoi construction
of the points which define a triangular lattice. This is because the
triangular lattice and the honeycomb structure together define a
Voronoi/Delaunay dual.

In each case the tessellation in the z-plane can be thought of as
consisting a vertex to which their are connected a number of straight
edges of finite length. In the following we shall examine the effect a
conformal mapping has on these edges. We shall compute their curvature
in the image plane - from which we can compute the mean and mean
square curvature of the edges connected to a given vertex in the image
plane.

\subsection{ The Complex Curvature of an Edge Upon Being Mapped to the
  Image Plane}

Upon applying an analytical mapping, the vertex will now be located at
a point f(p)=u+iv in the w-plane. Since all the edges are straight,
using Eq. ($\!\!$~\ref{eq:curvature}) it is found that the m edges
have a new curvature in the image plane given by,
\begin{equation}
  \widetilde{\kappa_n}
  =
  \frac{1}{|f_{1}(p)|}
  \text{Im}
  \left[
    \frac{f_{2}(p)}{f_{1}(p)}
    \widehat{\xi_n}
  \right]
  =
  \frac{|f_{2}(p)|}{|f_{1}(p)|^2}
  \text{Im}
  \left[ 
    e^{i\Theta}
    \widehat{\xi_n}
  \right],
  \label{eq:mcurvatures}
\end{equation}
where we have set (i.e.  ${\kappa}_n=0$ i.e. the curvature of the
edges in the z plane). We also have
\[
\Theta =\theta_2-\theta_1=Arg \left[ \frac{f_{2}(p)}{f_{1}(p)} \right],
\]
and Arg stands for the complex argument. From here on we shall adopt
the notation that
\[
\theta_n=Arg [f_{n}(p)].
\]
The nth tangent vector is given by,
\[
\widehat{\xi_n} = e^{i(\frac{2\pi n}{m}+\gamma)},
\]
where $\gamma$ is the angle made by the 0th tangent vector, in the
z-plane, with the x axis. This is shown schematically in Fig.
(\ref{initial}). Eq. ($\!\!$~\ref{eq:mcurvatures}) can be written as,
\begin{equation}
  \widetilde{\kappa_n}
  =
  \frac{|f_{2}(p)|}{|f_{1}(p)|^2}
  \sin\left(\Theta+\frac{2\pi n}{m} + \gamma\right)
  =
  \frac{|f_{2}(p)|}{|f_{1}(p)|^2}
  \sin\left(s+\frac{2\pi n}{m}\right),
\label{eq:identity}
\end{equation}
where $s=\gamma+\Theta$. 

\subsection{The Mean Complex Curvature of the Edges, Connected to a Given
  Vertex, in the Image Plane}

Let us define the mean curvature of the edges in the z-plane as,
\begin{equation}
  C_m
  =
  \frac{1}{m}
  \sum^{n=m-1}_{n=0}
  \kappa_n,
  \nonumber
\label{eq:original_mean_curvature}
\end{equation}
similarly, the mean curvature of the edges in the w-plane as,
\begin{equation}
\widetilde{C_m}
=
\frac{1}{m}
\sum^{n=m-1}_{n=0}
\widetilde{\kappa_n}
=
\frac{1}{m}
\frac{|f_{2}(p)|}{|f_{1}(p)|^2}
\;
\sum^{n=m-1}_{n=0}
\sin\left(s+\frac{2\pi n}{m}\right)
=0,
\label{eq:mean_curvature}
\end{equation}
where we have substituted Eq. ($\!\!$~\ref{eq:identity}) into Eq.
($\!\!$~\ref{eq:mean_curvature}). This result is rather trivial it 
merely expresses the fact that in the image plane, connected to each
vertex, there are pairs of opposed edges for which the sum of their
curvatures always vanish.

\subsection{The Mean Square Value of the Complex Curvature of the
  Edges, Connected to a Given Vertex, in the Image Plane}

Since the mean curvature always vanishes, the mean square curvature of
the edges in the z-plane is simply defined as,
\begin{equation}
  Q_m
  =
  \frac{1}{m}
  \sum^{n=m-1}_{n=0}
  {k_n}^2,
\end{equation}
and in the w-plane the mean square curvature is,
\begin{equation}
  \widetilde{Q_m}
  =
  \frac{1}{m}
  \sum^{n=m-1}_{n=0}
  \widetilde{k_n}^2
   =
  \frac{1}{m}
  \left(\frac{|f_{2}(p)|}{|f_{1}(p)|^2}\right)^2
  \;
  \sum^{n=m-1}_{n=0}
  \sin^2 
  \left( s+\frac{2\pi n}{m} \right)
  =
  \frac{1}{2}
  \left(\frac{|f_{2}(p)|}{|f_{1}(p)|^2}\right)^2.
\label{eq:mean_square_in_w_plane}
\end{equation}
where we have substituted Eq. ($\!\!$~\ref{eq:identity}) into Eq.
($\!\!$~\ref{eq:mean_square_in_w_plane}). Thus the mean square curvature of the edges, 
connected to a given vertex in the image plane, is independent with respect to the
orientation of the vertex in the image or z planes.

\begin{figure}
\begin{center}
\includegraphics[width=1.0\columnwidth]{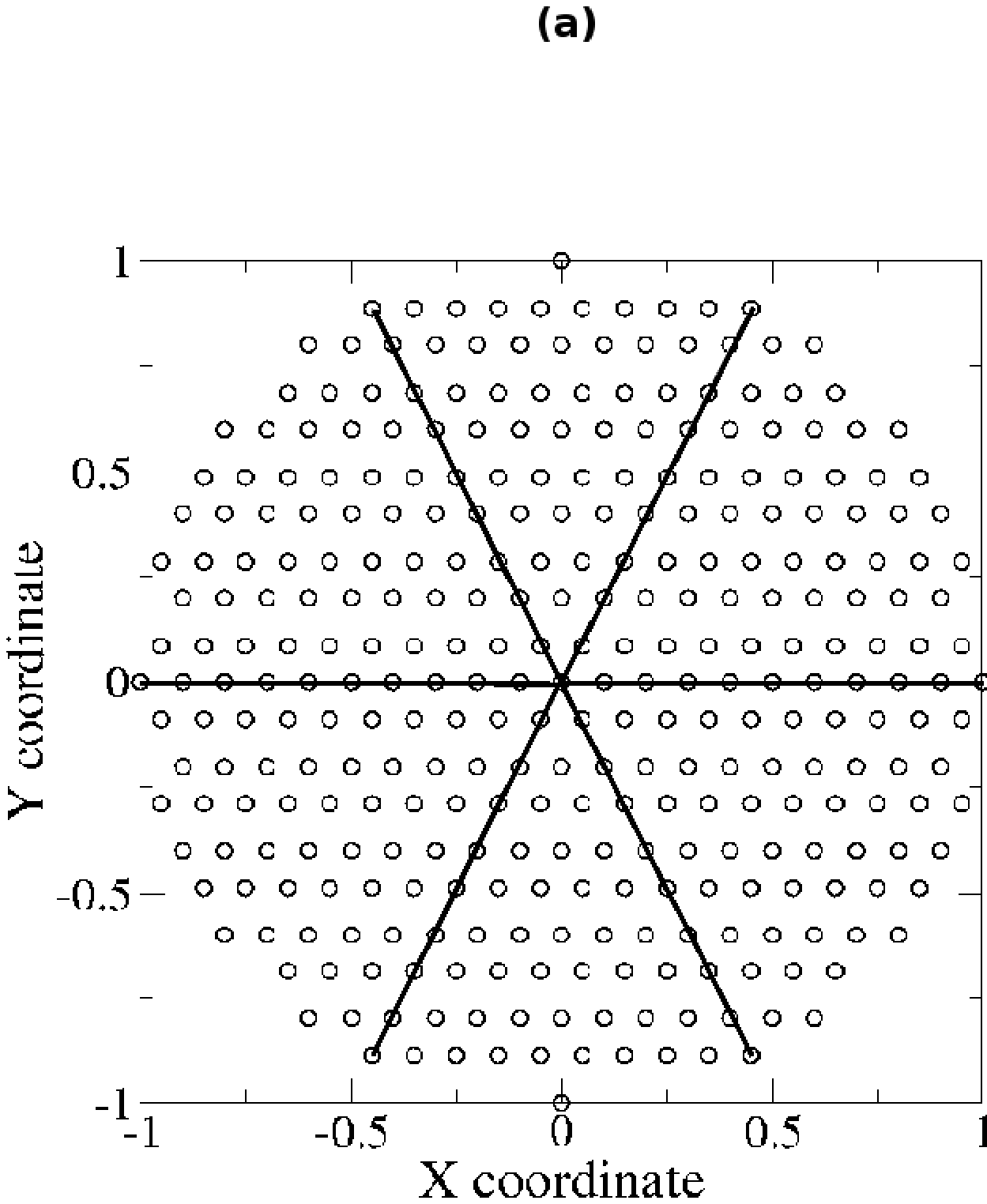}
\caption{{\bf (a)}: a circular disk cut out of a triangular
    lattice in the z plane, where the angle between each successive
    heavy line is $\pi/3$. Clearly the lattice lines are free of
    curvature. Upon applying a conformal transformation, {\bf (b)}: $w=z^{1/2}$ 
    and {\bf (c)}: complex inversion, i.e. $w=1/z$, the result is a conformal lattice in the image plane.
    Note that in the case of $w=z^{1/2}$ the conformal lattice has
    been produced by allowing the range of the z plane to extend to
    $\phi < 4\pi$.  In both cases the mean curvature vanishes
    everywhere but the mean square curvature does not. Also the mean
    square curvature diverges at the origin, which for both transformations
    is a critical point. At the critical point the angle
    preserving property of the transformation breaks down, this can be
    easily seen in the case of the $w=z^{1/2}$ transformation -
    instead of being preserved, the angle $\pi/3$ between rays (heavy
    lines) emanating from the origin is halved.}
\label{conformal_transformation}
\end{center}
\end{figure}

\subsection{Examples}

To illustrate the results derived for the mean and mean square
curvature, we now examine two conformal lattices which have been
generated by applying a transformation to a circular disk cut out of a
triangular lattice. Just such a cut out is shown in Fig.(\ref{conformal_transformation}a), it can be seen that the
lattice is free of curvature. Applying the mapping $f(z)=z^{1/2}$ to the 
triangular lattice in the z-plane yields a conformal lattice in
the image plane, see 
Fig.(\ref{conformal_transformation}b). From Eq. ($\!\!$~\ref{eq:mean_square_in_w_plane}) the 
mean square curvature is found to be
\begin{equation}
\widetilde{Q_m} = 
\frac{1}{2r}
=
\frac{1}{2R^2},
\label{eq:mean_square_analytical_1}
\end{equation}
where we have transformed from the coordinates in the z-plane
$(r,\lambda)$ to the w-plane coordinates $(R,\Lambda)$. Similarly, the effect of inversion is shown in the bottom part of
Fig.(\ref{conformal_transformation}). To compute the mean square
curvature we substitute $f(z)=1/z$ into Eq.  ($\!\!$~\ref{eq:mean_square_in_w_plane}), which yields,
\begin{equation}
  \widetilde{Q_m} = 
  \frac{r^2}{2}
  =
  \frac{2}{R^2}.
\label{eq:mean_square_analytical_2}
\end{equation}

\begin{figure}
\begin{center}
\includegraphics[width=0.3\columnwidth]{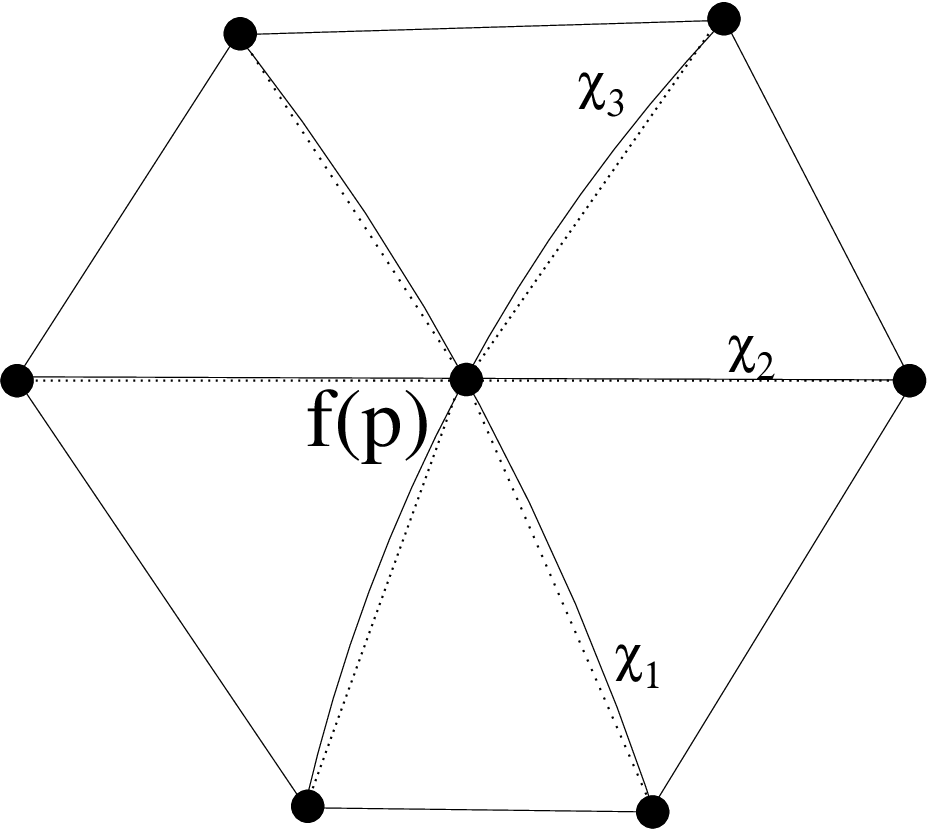}
\caption{This diagram shows the effect of the transformation f(z) on
  a hexagonal lattice cell, which in the z plane has its centre at p.
  In the w-plane the lattice cell becomes deformed and its centre is
  now to be found at f(p). The deformed hexagonal cell can be
  decomposed into three arcs which cross the centre of the cell. Each
  arc can in turn be decomposed into points, to which we fit the
  equation of a circle. From this it is possible to calculate the
  curvature of each arc, represented here by $\chi_1, \chi_2$ and
  $\chi_3$. }
\label{hexa_curve}
\end{center}
\end{figure}

\begin{figure}
\begin{center}
\includegraphics[width=1.0\columnwidth]{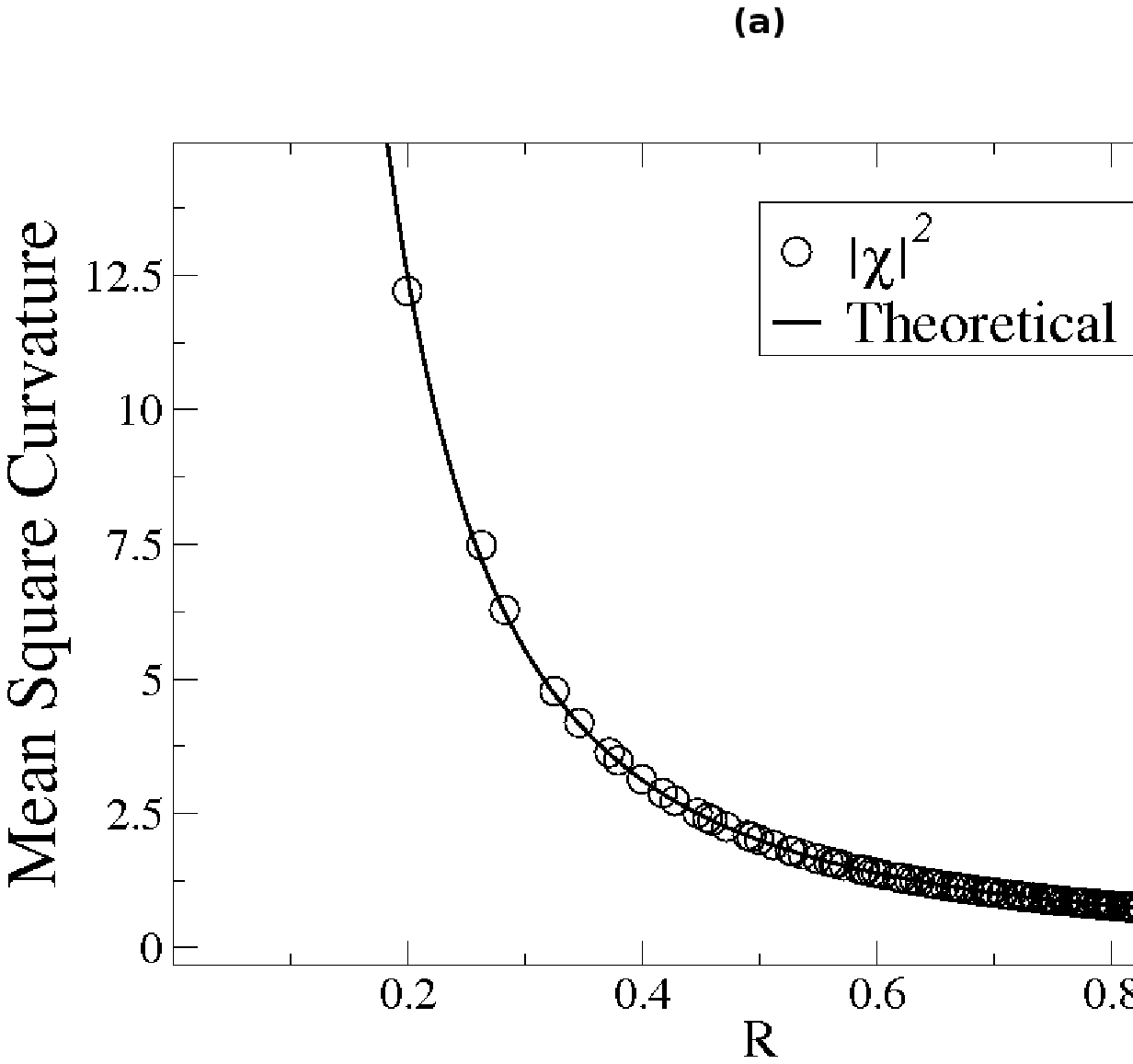}
\caption{ Mean square curvature for the mapping {\bf (a)} $f(z)=z^{1/2}$ and {\bf (b)} $f(z)=1/z$.}
\label{comp1}
\end{center}
\end{figure}

To verify our results we estimate the mean square curvature directly
from the transformed lattice. To do so, note that any given hexagonal
cell in the image plane can be decomposed into three arcs which cross
the the centre of the cell at f(p), as shown in Fig.
(\ref{hexa_curve}). Each of these three arcs can be further decomposed
into three points which can be used to fit the equation of a circle,
from which the curvature of the arc can be readily computed. Let us
denote the three directly computed curvatures as $\chi_1, \chi_2$ and
$\chi_3$. We can define the mean square curvature as,
\begin{equation}
|\chi|^2
=
\frac{1}{3}
(\chi_1^2+\chi_2^2+\chi_3^2).
\label{eq:numerical_msc}
\end{equation}
Thus, for the two conformal lattices shown in fig 4, we can estimate
the mean square curvature for a given vertex using Eq.
($\!\!$~\ref{eq:numerical_msc}). The results are shown by the circles in Fig. (\ref{comp1}) along with the
theoretical results Eq. ($\!\!$~\ref{eq:mean_square_analytical_1})
and Eq. ($\!\!$~\ref{eq:mean_square_analytical_2}). 

Except in the case of complex inversion there will be a difference
between the analytical value of the curvature and the estimated
curvature. This is due to the fact that the curves to which we have
fitted the circles are not themselves circular arcs. We quantify this
discrepancy in section VI of this paper. 

\section{Some Generalisations}

In the previous section we considered a vertex, in the z-plane, to
which there are connected a finite number of straight edges. We found
that when the vertex is mapped to the image plane the mean curvature
of the edges vanishes and that the mean square curvature is
independent with respect to the orientation of the vertex.

Let us now consider the case where the edges connected to a vertex, in
the z-plane, have some initial curvature. Let us also assume that the
number of edges connected to the vertex is infinite. As the number of
edges tends to infinity, the tangent vectors together describe a
circle of unit radius centred on the vertex. Upon averaging over the
length of the circle the mean curvature of the edges in the z-plane
can be written as,
\begin{equation}
  C
  =
  \frac{1}{2 \pi}
  \int^{2\pi}_{0}
  \kappa(\phi)
  d\phi,
\label{eq:mean_continuum}
\end{equation}
where $\kappa (\phi)$ is now a continuous function over the range $0
\leq \phi < 2\pi$ and describes the curvature of the edges in the
z-plane. Similarly, the mean curvature of the edges upon being
transformed to the w-plane is given by,
\begin{equation}
\widetilde{C}
=
\frac{1}{2 \pi}
\int^{2\pi}_{0}
\widetilde{\kappa(\phi)}
d\phi.
\label{eq:mean_continuum}
\end{equation}
where we define,
\begin{equation}
\widetilde{\kappa (\phi)}
=
\frac{1}{|f_{1}(p)|}
\text{Im}
\left[
\frac{f_{2}(p)}{f_{1}(p)}
\widehat{\xi (\phi)}
\right]
+
\frac{\kappa (\phi)}{|f_{1}(p)|},
\label{eq:curvature_cont1}
\end{equation}
Eq. ($\!\!$~\ref{eq:curvature_cont1}) can be written as
\begin{equation}
  \widetilde{\kappa (\phi)}  
  =
  \frac{1}{|f_{1}(p)|}
  \text{Im}
  \left[ 
    Te^{i\Theta}
    \widehat{\xi (\phi)}
  \right]
  +
  \frac{\kappa (\phi)}{|f_{1}(p)|}
  =
  \frac{|f_{2}(p)|}{|f_{1}(p)|^2}
  \text{Im}
  \left[ 
    e^{i\Theta}
    \widehat{\xi (\phi)}
  \right]
  +
  \frac{\kappa (\phi)}{|f_{1}(p)|},
  \nonumber
\end{equation}
where T and $\Theta$ have the same definitions as in section IV.
Writing out $\widehat{\xi (\phi)}$ explicitly and simplifying gives
\begin{equation}
\widetilde{\kappa (\phi)}
=
\frac{|f_{2}(p)|}{|f_{1}(p)|^2}
\sin(\Theta + \phi) +
\frac{\kappa (\phi)}{|f_{1}(p)|}.
\label{eq:curvature_cont2}
\end{equation}
Substituting Eq. ($\!\!$~\ref{eq:curvature_cont2}) into Eq.
($\!\!$~\ref{eq:mean_continuum}) yields,
\begin{equation}
\widetilde{C}
=
\frac{1}{2 \pi}
\frac{|f_{2}(p)|}{|f_{1}(p)|^2}
\int^{2\pi}_{0}
\sin(\Theta+\phi)
d\phi
+
\frac{C}{|f_{1}(p)|}
=
\frac{C}{|f_{1}(p)|}.
\end{equation}
Thus the conformal transformation scales the original mean curvature
by the magnification factor $1/|f_{1}(p)|$ of the transformation (see Eq.
($\!\!$~\ref{eq:lscale})).

Since the mean curvature does not necessarily vanish, we define the
mean square curvature in the z-plane as,
\begin{equation}
Q
=
\frac{1}{2\pi}
\int^{2\pi}_{0}
\left(
\kappa (\phi)
-
C
\right)^2
d\phi,
\label{eq:z_mean_square_cur_cont}
\end{equation}
similarly we define the mean square curvature in the w-plane as,
\begin{equation}
\widetilde{Q}
=
\frac{1}{2\pi}
\int^{2\pi}_{0}
\left(
\widetilde{\kappa (\phi)}
-
\widetilde{C}
\right)^2
d\phi
=
\frac{1}{2\pi}
\int^{2\pi}_{0}
\left(
\widetilde{\kappa (\phi)}^2
-
2\widetilde{C}\widetilde{\kappa (\phi)}
+
\widetilde{C}^2
\right)
d\phi,
\label{eq:w_mean_square_cur_cont_prev}
\end{equation}
substituting Eq. ($\!\!$~\ref{eq:curvature_cont2}) into Eq.
($\!\!$~\ref{eq:w_mean_square_cur_cont_prev}) and performing the
resulting integrals gives,
\begin{equation}
\widetilde{Q}
=
\frac{1}{2}
\left(
\frac{|f_{2}(p)|}{|f_{1}(p)|^2}
\right)^2
+
\frac{Q}{|f_{1}(p)|^2}
+
\frac{1}{\pi}
\frac{|f_{2}(p)|}{|f_{1}(p)|^3}
\int^{2\pi}_{0}
\kappa (\phi)
\sin (\Theta+\phi)
d\phi.
\label{eq:w_mean_square_final}
\end{equation}
The first term is the mean square curvature of the edges in the image
plane if the edges in the z-plane have no curvature. The second term
simply states that the original mean square curvature of the edges is
scaled by a factor of $1\/|f_{1}(p)|^2$. The third term is a coupling
between the original curvature of the edges and the transformation
$f(z)$. To get a better understanding of the third term let us label it as
$\widetilde{Q_3}$. It can be written as,
\begin{equation}
  \widetilde{Q_3}
  =
  \frac{1}{\pi}
  \frac{|f_{2}(p)|}{|f_{1}(p)|^3}
  \int^{2\pi}_{0}
  \kappa (\phi)
  (\sin\Theta\cos\phi + \cos\Theta\sin\phi)
  d\phi.
\label{eq:third_term}
\end{equation}
Since $\kappa (\phi)$ is an arbitrary function which exists over the
interval $0 \leq \phi < 2\pi$, it is natural to express it in terms of
a Fourier series, so that,
\begin{equation}
\kappa (\phi)=\frac{1}{2}a_0 + \sum^{\infty}_{q=1} a_q \cos(q\phi) + b_q \sin(q\phi).
\label{eq:fourier}
\end{equation}
Substituting Eq. ($\!\!$~\ref{eq:fourier}) into Eq.
($\!\!$~\ref{eq:third_term}) we find,
\begin{equation}
  \widetilde{Q_3}
  =
  \frac{|f_{2}(p)|}{|f_{1}(p)|^3}
  (a_1\sin\Theta + b_1\cos\Theta),
\label{eq:q3}
\end{equation}
thus the only contribution to the mean square curvature is from the
mode $q=1$. The contributions from all other modes cancel out, so that
an increase in mean square curvature for a given edge is cancelled out
by a decrease in mean square curvature for some other edge.
Furthermore Eq. ($\!\!$~\ref{eq:q3}) depends on the orientation of the
vertex in the z-plane through the angle $\Theta$. The turning points (i.e.
the the maximum and minimum mean square curvature) depend
on the values of the co-efficients $a_1$ and $b_1$ and are located at
$\Theta=tan^{-1}(a_1/b_1)$ and $\Theta=tan^{-1}(a_1/b_1)+ \pi$. 

Thus when the edges in the z-plane are not free of curvature the
effect of the conformal mapping is to scale the mean curvature, while
the mean square curvature is found to be no longer independent with
respect to orientation of the vertex.

\section{Higher Order Terms in the Complex Curvature}

\begin{figure}
\begin{center}
\includegraphics[width=0.7\columnwidth]{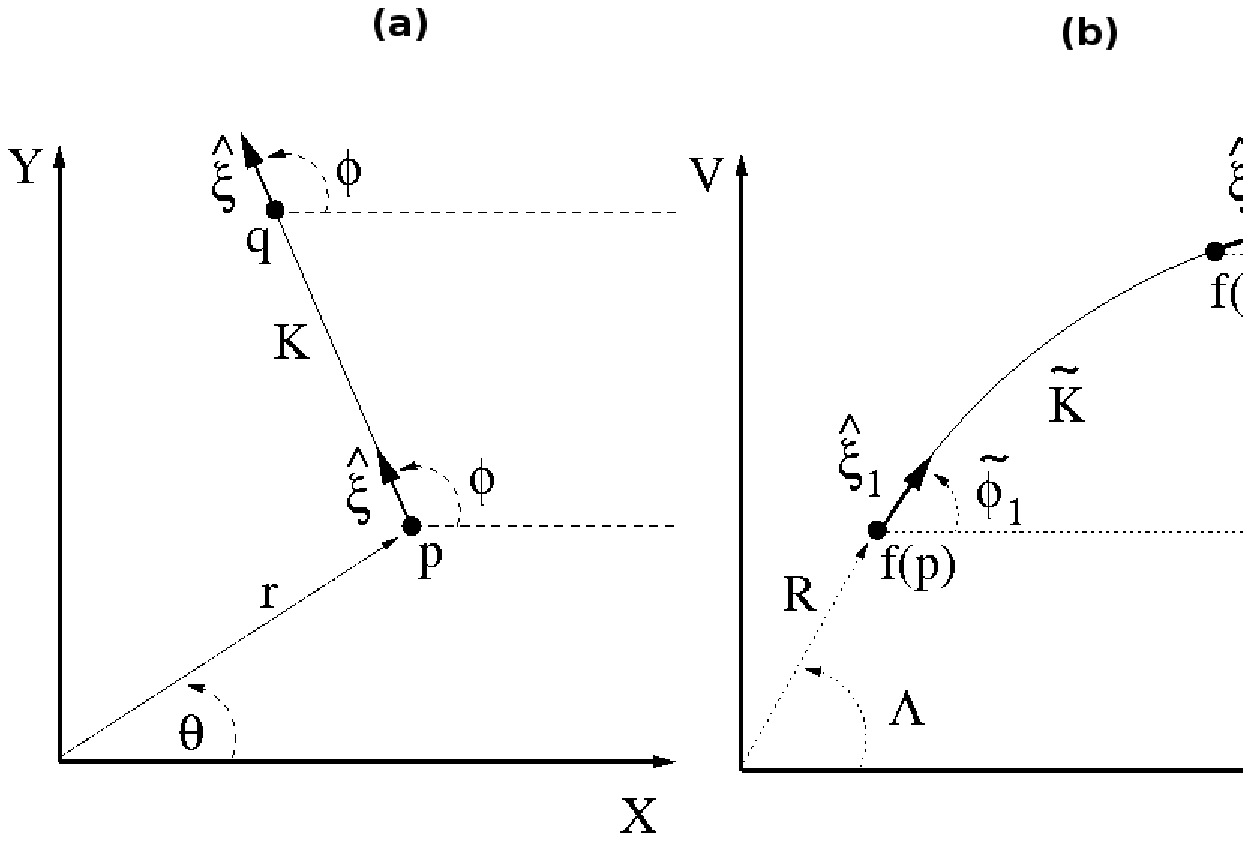}
\caption{{\bf (a)} This diagram shows a straight line K in the
  z-plane. The line starts at the point labelled p and ends at q, it
  has a total length $L$. At both points the unit tangent vector has
  the same direction and makes an angle of $\phi$ with the x-axis.
  {\bf (b)} This diagram shows the resulting curve $\widetilde
  {\mathrm K}$ in the image plane due to the conformal mapping
  $w=f(z)$. The curve begins at the point labelled $f(p)$ and ends at
  $f(q)$ and now has a length $\widetilde l$. In general the angle
  that the tangent vectors make at $f(p)$ and $f(q)$, which are
  labelled $\widetilde{\phi_1}$ and $\widetilde{\phi_2}$ respectively,
  will no longer be equal.}
\label{curvature1}
\end{center}
\end{figure}

For a given 2D plane curve its curvature is defined as the rate of
change of the angle of the tangent vector with respect to the distance
along the curve. We can think of this as limiting process: take two
tangent vectors a short distance apart on the curve, measure the angle
they make with respect to a fixed axis, and take the limit in which
the separation between the tangent vectors vanishes. This then gives
the instantaneous value of the curvature.

Consider a 2D plane curve in the image plane which has been generated
by applying a complex mapping to some curve in the z plane. We can
choose a pair of points on the image curve a {\it finite} distance
apart and for both points construct a tangent vector. Thus we can ask:
what is the rule which gives the total difference in angle between the
two tangent vectors?

In this section we derive an expression for the {\it average complex
curvature} which we define as: the total change in the angle of the
tangent vector over a segment of a curve averaged over the length of
the segment. The result is a series expansion in which Eq.
($\!\!$~\ref{eq:curvature}) is the lowest order term. This series
expansion will lead us to an expression which can be used to measure
the degree to which a given image arc differs from being perfectly
circular. Here we shall only consider the effect of the mapping $w=f(z)$ on a
straight line in the z-plane. Furthermore, we shall assume that the
effect of the mapping is to produce a open, simple and well behaved
curve in the image plane.

Consider a straight line K in the z plane, which joins point $p$ to
$q$ and let us assume it has a length $|\xi|$. This is shown in Fig. (\ref{curvature1}a). At $p$ we have drawn the unit
tangent vector $\widehat{\xi}=e^{i\phi}$, so that the vector joining
$p$ to $q$ is given by
\begin{equation}
\xi=|\xi|\widehat{\xi(\phi)}.
\label{eq:myvector}
\end{equation}
The line K is described by the equation
\begin{equation}
{\mathrm K}=z(t)=p+t\widehat{\xi}=re^{i\theta}+te^{i\phi},
\label{eq:zequ}
\end{equation}
where $0 \leq t \leq |\xi|$. Note that at $t=0$ we have $z=p$ and at
$t=|\xi|$ we have 
\[
z(|\xi|)=q=p+\xi=re^{i\theta}+|\xi|e^{i\phi}.
\]
Thus K has been parametrised by the free parameter t which measures
the distance along K from the point $p$. The line K has a real component and imaginary components 
\[
x(t)=r\cos\theta+t\cos\phi \;\;\; \textrm{and} \;\;\; y(t)=r\sin\theta+t\sin\phi.
\]
respectively.  

Now consider what happens if we apply an analytical mapping $w=f(z)$
to K, the result is a new curve $\widetilde {\mathrm K}$ in the image
plane which starts at the point labelled $f(p)$ and ends at $f(q)$ -
as shown in Fig. (\ref{curvature1}b). This curve now
has some new length which we call $l$ (not to be confused with the
chord distance between $f(p)$ and $f(q)$). We define the {\it average} complex curvature of the image curve
$\widetilde {\mathrm K}$ as
\begin{equation}
\widetilde{\mathbb{K}}
=
\frac{\Delta\widetilde{\phi}}{l}
=\frac{\widetilde{\phi_2}-\widetilde{\phi_1}}{l}.
\label{eq:averageK}
\end{equation}
where the tangent vectors at $f(p)$ and $f(q)$ are labelled
${\widehat{\xi_1}}$ and ${\widehat{\xi_2}}$ and make an angle of
$\widetilde{\phi_1}$ and $\widetilde{\phi_2}$, respectively, with the
the u axis. 

If for Eq. ($\!\!$~\ref{eq:averageK}) we keep the point
$p$ fixed and take the limit in which $|\xi|$ goes to zero we recover
the instantaneous curvature of the curve $\widetilde {\mathrm K}$ at
the point $f(p)$, that is
\[
\widetilde{\kappa}(p)=\lim_{|\xi|\to 0} \widetilde{\mathbb{K}},
\]  

Let us now define the difference between the instantaneous curvature
of the image arc at its starting point and the average curvature of the
arc, which we denote as
\[
\Delta\widetilde{\mathbb{K}}=
\widetilde{\mathbb{K}}
-
{\widetilde \kappa}(p).
\]
We note that if the image arc is a perfect circle then the
instantaneous curvature has a constant value at every point on the arc
and therefore $|\Delta\widetilde{\mathbb{K}}|=0$. This fact can be
used to quantify the degree to which the image curve deviates from
perfect circularity.

In the context of 2D and quasi 2D foams it is important to know the
degree to which a given arc, generated by a conformal mapping,
deviates from perfect circularity. Laplace's law requires that the
curvature of a soap film is constant, therefore for a 2D soap film a
large value of $|\Delta\widetilde{\mathbb{K}}|$ implies a large error
in how well the soap film approximates the conformal lattice. The same
holds true for quasi 2D foams (in the constant volume regime), except
that we have to remember that the total curvature of the soap film,
i.e. the sum of longitudinal ($\kappa=\kappa(p)$)and transverse
($\kappa_t$) curvatures, has an instantaneous value of
\[
H=-\frac{1}{2}{\widetilde \kappa}(p).
\]
In either case Laplace's law is satisfied only if the soap film has a
constant longitudinal curvature, i.e. it describes a circular arc when
viewed from a direction perpendicular to the bottom plate of the
Hele-Shaw cell.

In the following we shall decompose the task of finding
$\widetilde{\mathbb{K}}$ into two parts: firstly we compute $l$ and
secondly we compute $\Delta\widetilde{\phi}$.

\subsection{Arc Length in the Image Plane}

As stated above: the real and imaginary components of the line K are a
function of the free parameter $t$. Upon mapping K to the image plane,
(using the analytical function $w=f(z)$) the result is the image curve
$\widetilde{{\mathrm K}}$ - the real and imaginary components of which
are also functions of the free parameter $t$ and denoted by $u(t)$ and
$v(t)$, respectively. We can write the real and imaginary components
of $\widetilde{{\mathrm K}}$ in the form of a series expansion about
the point $p$ (i.e. $t=0$), giving
\begin{equation}
  u(t)=u+t\frac{u_{t}}{1!}+{t^2}\frac{u_{tt}}{2!}+...+\mathrm{h.o.t.}
\label{eq:uexp}
\end{equation}
and also
\begin{equation}
  v(t)=v+t\frac{v_{t}}{1!}+{t^2}\frac{v_{tt}}{2!}+...+\mathrm{h.o.t.}
\label{eq:vexp}
\end{equation}
where,
\begin{equation}
u=\left.u(t)\right|_{t=0}=u(0)
\label{eq:1div}
\end{equation}
\begin{equation}
u_t=\left.\frac{du(t)}{dt}\right|_{t=0}
\label{eq:2div}
\end{equation}
\begin{equation}
u_{tt}=\left.\frac{d^2u(t)}{dt^2}\right|_{t=0}
\label{eq:3div}
\end{equation}
and so on. We also define $v_{t}$ and $v_{tt}$ (and higher order
terms) in a similar manner.

The total length of the arc $\widetilde {\mathrm K}$ is given by,
\begin{equation}
l
=
\int^{|\xi|}_{0}
\sqrt{
\left( \frac{du(t)}{dt}\right)^2
+
\left( \frac{dv(t)}{dt}\right)^2
}
dt.
\label{eq:arclength}
\end{equation}
We can substitute Eq.($\!\!$~\ref{eq:uexp}) and
Eq.($\!\!$~\ref{eq:vexp}) into Eq.($\!\!$~\ref{eq:arclength}), then
expand the integrand in powers of $t$, and integrate each term
with respect to $t$ to give,
\begin{equation}
  l = l_0{|\xi|}+ l_1|\xi|^2+l_2|\xi|^3...+\mathrm{h.o.t.}
\label{eq:lexp}
\end{equation}
where we assume that $|\xi|$ is small enough to guarantee convergence
of the series. The coefficients in Eq.($\!\!$~\ref{eq:lexp}) are
defined as
\begin{equation}
l_0=\sqrt{{u_{t}}^2+{v_{t}}^2}
\label{eq:l0}
\end{equation}
\begin{equation}
l_1=\frac{1}{2}\frac{u_{t}u_{tt}+v_{t}v_{tt}}{\sqrt{{u_{t}}^2+{v_{t}}^2}}
\label{eq:l1}
\end{equation}
\begin{equation}
l_2
=
\frac{1}{6}\sqrt{{u_{t}}^2+{v_{t}}^2}
\left[
\frac{{u_{tt}}^2 + {u_{t}}{u_{ttt}}+{v_{t}}{v_{ttt}} + {v_{tt}}^2}{\sqrt{{u_{t}}^2+{v_{t}}^2}}
-
\frac{(u_{t}u_{tt}+v_{t}v_{tt})^2}{2({u_{t}}^2+{v_{t}}^2)^2}
\right].
\label{eq:l2}
\end{equation}
We notice that the coefficients given by Eq.($\!\!$~\ref{eq:l0}) to
Eq.($\!\!$~\ref{eq:l2}) (and higher order terms) are defined with
respect to the parameter $t$.  In order to express the coefficients in
terms of the complex number $z$ we use Eq.  ($\!\!$~\ref{eq:zequ})
from which we have the relationship $dz/dt=e^{i\phi}$, this can be
rearranged to give $dt=e^{-i\phi}dz$. Thus Eq.($\!\!$~\ref{eq:1div}) to Eq.($\!\!$~\ref{eq:3div}) (and
higher order derivatives) can be written as
\begin{eqnarray}
\left.
\frac{d^nu}{dt^n} 
\right|_{t=0}
&=&
{\mathrm Re}\left[\left. \frac{d^nw}{dt^n} \right|_{t=0} \right] 
\nonumber
=
{\mathrm Re}\left[\left. \frac{d^nw}{(e^{-i\phi}dz)^n} \right|_{z=p} \right] 
=
{\mathrm Re}\left[\left. \frac{d^nw}{dz^n} \right|_{z=p} e^{in\phi}\right] 
\nonumber
\\
&=&
|f_n(p)| {\mathrm Re}\left[e^{i(\theta_n+n\phi)}\right] 
=
|f_{n}(p)|\cos(\theta_n +n\phi)
\label{eq:realt}
\end{eqnarray}
where we have made a
change of variable from $t$ to $z$ and used the fact that at $t=0$ we
have $z=p$; also we used 
\[
\left.
\frac{d^nw}{dz^n}
\right|_{z=p}
=
\left.
|f_{n}(z)|
e^{iArg[f_n(z)]}
\right|_{z=p}
=
|f_{n}(p)|
e^{i\theta_n}.
\]
Similarly for the derivatives $v_t$ and $v_{tt}$ (and higher order
terms), we have,
\begin{equation}
\left.
\frac{d^nv}{dt^n} 
\right|_{t=0}
=
|f_{n}(p)|\sin(\theta_n +n\phi).
\label{eq:imaginaryt}
\end{equation}
Finally, the coefficients given by Eq.($\!\!$~\ref{eq:l0}) to
Eq.($\!\!$~\ref{eq:l2}) can be expressed in terms of
Eq.($\!\!$~\ref{eq:realt}) and Eq.($\!\!$~\ref{eq:imaginaryt}) to give
\begin{equation}
l_0=|f_{1}(p)|
\label{eq:fl0}
\end{equation}
\begin{equation}
l_1
=
\frac{|f_{1}(p)|}{2}\mathrm{Re}\left[\frac{f_2(p)}{f_1(p)}\widehat{\xi(\phi)}\right]
\label{eq:fl1}
\end{equation}
\begin{equation}
l_2
=
\frac{|f_{1}(p)|}{12}
\left(
\frac{|f_{2}(p)|^2}{|f_{1}(p)|^2}
-
\mathrm{Re}
\left[
\left(
\frac{f_2(p)}{f_1(p)}\widehat{\xi(\phi)}
\right)^2
-
2
\left(
\frac{f_3(p)}{f_1(p)}\widehat{\xi(\phi)}
^2
\right)
\right]
\right)
\label{eq:fl2}
\end{equation}

Upon substituting Eq.($\!\!$~\ref{eq:fl0}), Eq.($\!\!$~\ref{eq:fl1})
and Eq.($\!\!$~\ref{eq:fl2}) into Eq.($\!\!$~\ref{eq:lexp}) we have a
series expansion in powers of the $|\xi|$. We note that the lowest
order term in Eq.($\!\!$~\ref{eq:lexp}) is identical to
Eq.($\!\!$~\ref{eq:lscale}).

\subsection{Change in the angle of the tangent vector in the image plane}

\begin{figure}
\begin{center}
\includegraphics[width=1.0\columnwidth]{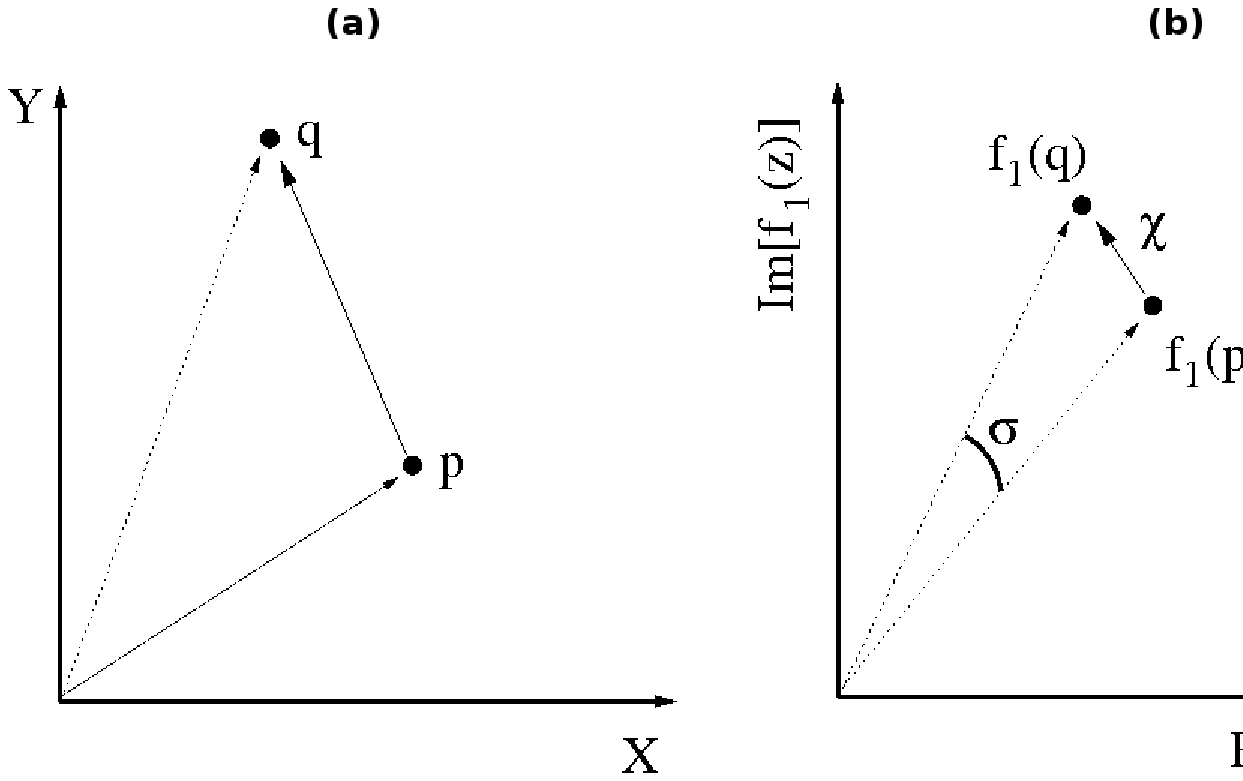}
\caption{After Needham (1997): {\bf (a)} This diagram shows two points p and q in the z
  plane. {\bf (b)} Upon applying the mapping $f_1(z)=df(z)/dz$ the
  points are mapped to $f_1(p)$ and $f_1(q)$, which are connected by
  the vector $\chi$. Note that to transform $f_1(p)$ into $f_1(q)$ it
  is necessary to magnify $f_1(p)$ by some factor and to rotate
  $f_1(p)$ by the angle denoted by $\sigma$. {\bf (c)} This triangle is the result of dividing all the vectors shown
  in (b) by the complex
  number $f_1(p)$. Dividing by $f_1(p)$ has two effects: firstly the
  triangle is rotated so that one of its sides is now parallel to the
  real axis, secondly all the sides of the triangle are uniformly
  scaled.  This second effect means that the new triangle is similar
  to the triangle shown (b).}
\label{curvature2}
\end{center}
\end{figure}

We now compute the difference in angle,
$\Delta\widetilde{\phi}=\widetilde{\phi_2}-\widetilde{\phi_1}$, of the
tangent vectors in the image plane. This derivation is adapted from
the one used by Needham (1997) to compute the instantaneous curvature. The difference here is that we endeavour to retain all
details which may yield higher order terms in the expression for
$\Delta\widetilde{\phi}$.

Consider again the the straight line K in the z-plane, as shown Fig.
(\ref{curvature1}). Note that the tangent vector at $p$ upon being
mapped by the function $w=f(z)$ is, after being translated to $f(p)$,
magnified by a length $|f_{1}(p)|$ and rotated by an angle
$Arg[f_{1}(p)]$. Similarly, upon applying the mapping $w=f(z)$, the
tangent vector at $q$ is, after being translated to $f(q)$, magnified
by a length $|f_{1}(q)|$ and rotated by an angle $Arg[f_{1}(q)]$.
However, the rotation of the tangent vector at $f(q)$ will differ very
slightly by $\Delta\widetilde{\phi}$ from the rotation suffered by the
tangent vector at $f(p)$.

To compute the difference in angle between the tangent vectors
consider the two points p and q as shown in Fig.
(\ref{curvature2}a), upon applying the mapping $f_{1}(z)$ the points
are now located at $f_{1}(p)$ and $f_{1}(q)$, respectively. Where the
vector connecting $f_{1}(p)$ to $f_{1}(q)$ is labelled $\chi$ and is
given by
\begin{equation}
\chi
=
f_{1}(q)-f_{1}(p)
=
f_{2}(p)\widehat{\xi}|\xi|
+
\frac{1}{2!}f_{3}(p){\widehat{\xi}}^2|\xi|^2
+
\frac{1}{3!}f_{4}(p){\widehat{\xi}}^3|\xi|^3
...+\mathrm{h.o.t}
\nonumber
\label{eq:mychi}
\end{equation}
where we have made use of Eq.($\!\!$~\ref{eq:myvector}).

Note, with reference to Fig.
(\ref{curvature2}b), that to transform the vector $f_{1}(p)$ into the
vector $f_{1}(q)$ it is necessary to expand the length of $f_{1}(p)$
by some factor and to rotate the vector $f_{1}(p)$ by an angle
$\sigma$.  It is this angle $\sigma$ which is the extra rotation
suffered by the tangent vector at $f(q)$ compared to the rotation of
the tangent vector at $f(p)$, i.e. $\sigma=\Delta\widetilde{\phi}$. Our task 
is to compute the angle $\sigma$ shown on the right hand side
of Fig. (\ref{curvature2}). Note, $\sigma$ is part of a triangle whose
sides are given by the known vectors $\chi$, $f_{1}(p)$ and
$f_{1}(q)$. To compute $\sigma$ we rotate the triangle shown in Fig. (\ref{curvature2}b) to the real axis by
dividing by $f_{1}(p)$ (this is the same trick used by Needham (1997)), the result is shown in Fig. (\ref{curvature2}c),
this new triangle is geometrically similar to the original triangle
shown in Fig. (\ref{curvature2}b). The side
along the real axis has a length 1 and in addition the point $\nu$ is
given by the vector
\begin{equation}
\nu=[\chi/f_1(p)].
\label{eq:mynu} 
\end{equation}
By a simple application of trigonometry we have
\begin{equation}
\Delta\widetilde{\phi}=\sigma={\mathrm \tan}^{-1}\left({\frac{{\mathrm Im}[\nu]}{1+{\mathrm Re}[\nu]}}\right).
\label{eq:myphi}
\end{equation}
Upon substituting Eq.($\!\!$~\ref{eq:mynu}) into
Eq.($\!\!$~\ref{eq:myphi}), we can expand Eq.($\!\!$~\ref{eq:myphi})
in powers of $|\xi|$ (we assume that $|\xi|$ is small enough to allow
convergence) to give,
\begin{equation}
\Delta\widetilde{\phi}=
\sigma = \sigma_0{|\xi|}+
\sigma_1|\xi|^2+\sigma_2|\xi|^3...+\mathrm{h.o.t.}
\label{eq:sigma}
\end{equation}
where we have made use of Eq.($\!\!$~\ref{eq:mychi}), we find
\begin{equation}
\sigma_0
=
\mathrm{Im}
\left[
\frac{f_2(p)}{f_1(p)}
\widehat{\xi(\phi)}
\right]
\end{equation}
\begin{equation}
\sigma_1
=
\frac{1}{2}
\mathrm{Im}
\left[
\left(
\frac{f_3(p)}{f_1(p)}
-
\left(
\frac{f_2(p)}{f_1(p)}
\right)^2
\right)
\widehat{\xi(\phi)}^2
\right]
\end{equation}
\begin{equation}
\sigma_2
=
\mathrm{Im}
\left[
\left(
\frac{1}{3}
\left(
\frac{f_2(p)}{f_1(p)}
\right)^3
-
\frac{1}{2}
\left(
\frac{f_2(p)f_3(p)}{(f_1(p))^2}
\right)
+
\frac{1}{6}
\left(
\frac{f_4(p)}{f_1(p)}
\right)
\right)
\widehat{\xi(\phi)}^3
\right]
\end{equation}

\subsection{The Average Complex Curvature }

Substituting Eq.($\!\!$~\ref{eq:lexp}) and
Eq.($\!\!$~\ref{eq:sigma}) into Eq. ($\!\!$~\ref{eq:averageK}) and
expanding in powers of $|\xi|$ yields,
\begin{equation}
\widetilde{\mathbb{K}}
=
\widetilde{\mathbb{K}_0}
+
\widetilde{\mathbb{K}_1}
|\xi|
+
\widetilde{\mathbb{K}_2}
|\xi|^2
+
...\mathrm{h.o.t}
\label{eq:averageKexp}
\end{equation}
where we find,
\[
\widetilde{\mathbb{K}_0}
=
\widetilde{\kappa}
=
\frac{1}{|f_{1}(p)|}
\text{Im}
\left[
\frac{f_{2}(p)}{f_{1}(p)}
\widehat{\xi (\phi)}
\right]
\]
\begin{equation}
\widetilde{\mathbb{K}_1}
=
\frac{1}{2|f_{1}(p)|}
\left(
\text{Im}
\left[
\frac{f_{3}(p)}{f_{1}(p)}
\widehat{\xi (\phi)^2}
\right]
-
\frac{3}{2}
\text{Im}
\left[
\left(
\frac{f_{2}(p)}{f_{1}(p)}
\widehat{\xi (\phi)}
\right)^2
\right]
\right)
\nonumber
\end{equation}

\begin{eqnarray}
\widetilde{\mathbb{K}_2}
&=&
\frac{1}{48|f_{1}(p)|}
\left(
\frac{3|f_{2}(p)|^2}{|f_{1}(p)|^2}
\text{Im}
\left[
\frac{f_{2}(p)}{f_{1}(p)}
\widehat{\xi (\phi)}
\right]
+
27
\text{Im}
\left[
\left(
\frac{f_{2}(p)}{f_{1}(p)}
\widehat{\xi (\phi)}
\right)^3
\right]
\right.
\nonumber
\\
&-&
\frac{2|f_{2}(p)|^2}{|f_{1}(p)|^2}
\text{Im}
\left[
\frac{f_{3}(p)}{f_{2}(p)}
\widehat{\xi (\phi)}
\right]
-
34
\text{Im}
\left[
\frac{f_{2}(p)f_{3}(p)}{(f_{1}(p))^2}
\widehat{\xi (\phi)}^3
\right]
+
\left.
8
\text{Im}
\left[
\frac{f_{4}(p)}{f_{1}(p)}
\widehat{\xi (\phi)}^3
\right]
\right)
\nonumber
\end{eqnarray}

In effect Eq.($\!\!$~\ref{eq:averageKexp}) gives the total change in
the angle of the tangent vector in the image plane, as it traverses
the image curve, {\it averaged} over the length of the curve. We
observe that as $|\xi|\rightarrow0$ then
$\widetilde{\mathbb{K}}\rightarrow\widetilde{\mathbb{K}_0}=\widetilde{\kappa}$,
giving the instantaneous curvature of the image curve at the point
$f(p)$. 

\subsubsection*{Example $f(z)=z^{3}$}

\begin{figure}
\begin{center}
\includegraphics[width=0.9\columnwidth]{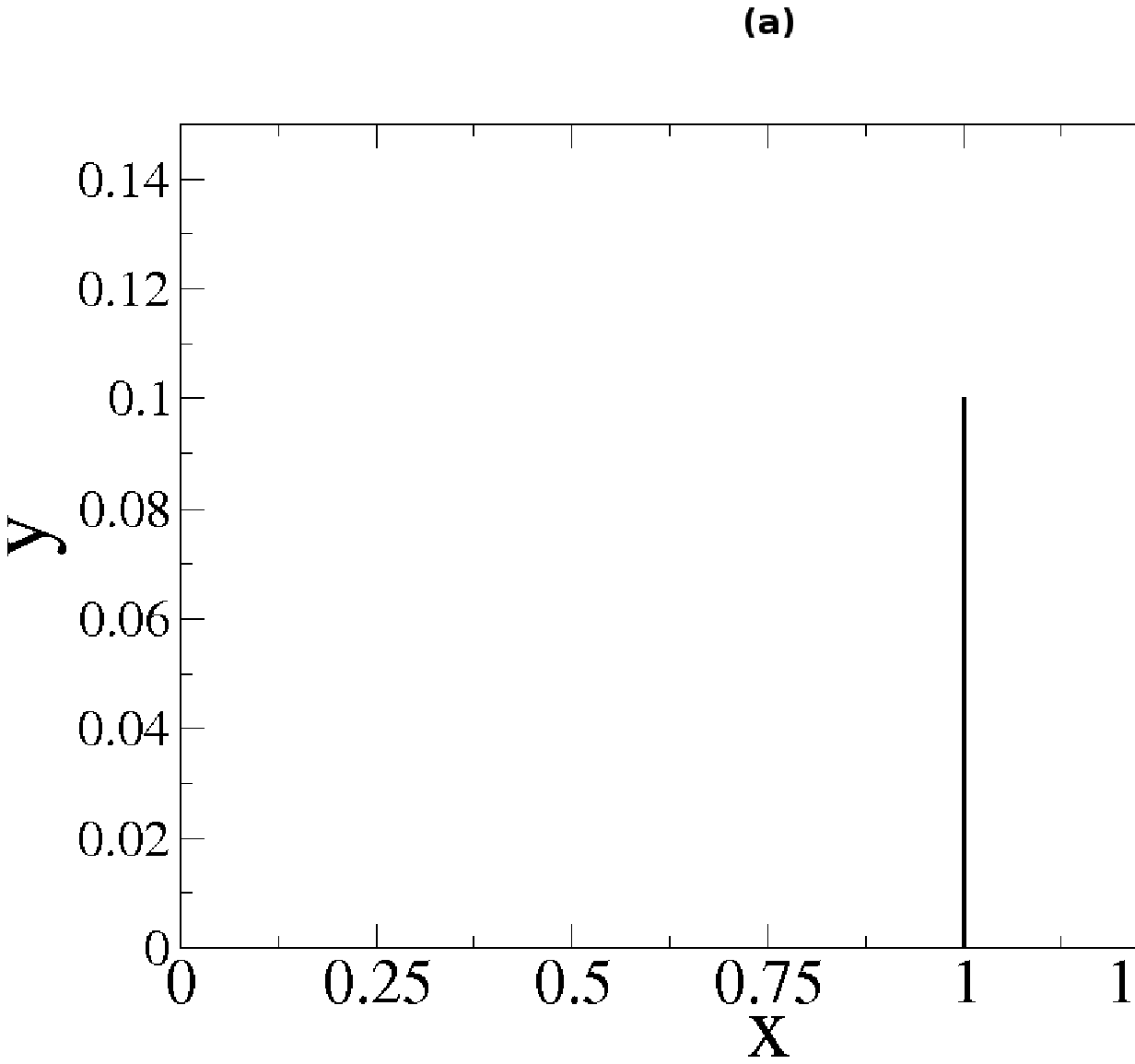}
\caption{{\bf (a)} A straight line in the z-plane, with
  $r=1,\theta=0,\phi=\pi/2$ and $|\xi|=0.1$ {\bf (b)} The resulting curve in the image plane generated by applying
  the conformal mapping $f(z)=z^3$ to the straight line defined by
  Eq.($\!\!$~\ref{eq:sample_straight_line})}
\label{straight_line}
\end{center}
\end{figure}

Let us apply Eq.($\!\!$~\ref{eq:averageKexp}) to a simple example,
we choose the complex mapping $f(z)=z^3$. Consider again the a straight line in the z-plane given by
Eq.($\!\!$~\ref{eq:zequ}), and let us set $r=1,\theta=0,\phi=\pi/2$
and $|\xi|=0.1$, so that we have
\begin{equation}
K=z(t)=1+te^{i\frac{\pi}{2}},
\label{eq:sample_straight_line}
\end{equation}
and $0 \leq t \leq 0.1$, the result is the straight line shown in Fig.
(\ref{straight_line}a). Upon applying the conformal transformation
$f(z)=z^3$ the result is the curve shown in Fig. (\ref{straight_line}b).

The derivatives for the function $f(z)=z^3$ are given by
$f_{1}(z)=3z^2$, $f_{2}(z)=6z$, $f_{3}(z)=6$, with all higher
derivatives being equal to zero. Upon evaluating these derivatives at
the point $p=re^{i\theta}=1e^{i0}=1$ we have $f_{1}(p)=3$,
$f_{2}(p)=6$ and $f_{3}(p)=6$; together with the angle of the tangent
vector to the straight line at the point p (i.e.  $\phi=\pi/2$) and
the length of the straight line in z-plane ($|\xi|=0.1$), we find
\[
\widetilde{\mathbb{K}_0}
=\frac{1}{3}
\mathrm{Im}
\left[
\frac{6}{3}
e^{i\pi/2}
\right]
=
\frac{2}{3}
\sin{\pi/2}
=
\frac{2}{3}
\]
and similarly $\widetilde{\mathbb{K}_1}=0$,
$\widetilde{\mathbb{K}_2} =-1/2.25$, $\widetilde{\mathbb{K}_3}=0$
and $\widetilde{\mathbb{K}_4}=0.281481$; thus up to fourth order in
$|\xi|$ we have $\widetilde{\mathbb{K}}\approx0.66225037$.

This value can be compared to a numerically computed value. To do so,
for the curve shown Fig. (\ref{straight_line}b), we need to compute the
length of the curve and the total change in the angle of the tangent
vector. This can be done by taking a large number of equally spaced
points on the curve. By drawing a straight line between any two
adjacent points we get a series of segments which approximate the
curve, the approximation improves as more points are taken. It is then
a simple matter for a computer to calculate the length of each segment
and to find the total length of the curve by summing up the lengths of
all the segments. To compute the total change in the angle of the
tangent vector: we first compute the angle that the first segment
makes with the u axis, then we compute the angle that the final
segment makes with the u axis, by taking the difference between the
two we find the change in the angle of the tangent vector between the
start and the end of the curve. Upon taking $1\textrm{x}10^5$ such
equally spaced points on the curve shown in Fig. (\ref{straight_line}b)
we have an numerically computed value for the average curvature,
giving $\widetilde{\mathbb{K}}_{{\mathrm{Num}}}=0.6622435$. It is
clear that Eq.($\!\!$~\ref{eq:averageKexp}) is converging to
$\widetilde{\mathbb{K}}_{{\mathrm{Num}}}$ as more terms are taken.

\subsection{Mean Square of the Average Complex Curvature}

For the instantaneous curvature we found it useful to compute the mean
square curvature, see Eq. ($\!\!$~\ref{eq:mean_square_in_w_plane}), which is independent of the orientation of the vertex and therefore serves as a useful measure of how strongly the conformal lattice is curved at any given point in the
image plane. Similarly we can compute a mean square value of the average complex curvature.
We find that this leads to a series expansion in which the lowest
order term is the mean square curvature given by Eq.
($\!\!$~\ref{eq:mean_square_in_w_plane}). When the higher order
terms in this expansion vanish it means that all the image edges, at
that particular point in the image plane, have a constant curvature.
If on the other hand the higher order terms do not vanish this means
that Laplace's law, for a 2D or quasi 2D foam used to realise the
conformal lattice, cannot be perfectly satisfied for every edge. The
magnitude of the higher order correction serve as an estimate of the
discrepancy between the foam and the conformal map, in most practical
cases the discrepancy can be assumed to be negligible if the leading
order correction in the expansion is small.

Consider again a vertex p in the z-plane to which there are connected
a number of straight edges of length $|\xi|$ separated by a constant
angular separation, such as that shown in Fig. (\ref{initial}). Upon
being mapped to the image plane the point p is now located at $f(p)$
and the straight edges are transformed into arcs. If we were to sum up
the average complex curvature, for each of the arcs emanating form the vertex
at $f(p)$, we would find that it vanishes.

If, however, for each arc we square the average complex curvature and sum up
the squares we have a quantity that does not vanish. Thus we can
compute the mean square value of the average complex curvature
$\widetilde{\mathbb{K}}$, which we shall denote by
$\widetilde{\mathbb{Q}}$, this can be done, most easily, by
integrating Eq.($\!\!$~\ref{eq:averageKexp}) with respect to $\phi$
over the range $0\leq \phi \leq 2\pi$ giving
\begin{equation}
\widetilde{\mathbb{Q}}
=
\frac{1}{2\pi}
\int^{2\pi}_{0}
\widetilde{\mathbb{K} (\phi)}
^2
d\phi.
\nonumber
=
\widetilde{\mathbb{Q}_0}
+
\widetilde{\mathbb{Q}_1}|\xi|^2
+
O(|\xi|^4)
\label{eq:w_mean_square_cur_cont}
\end{equation}
where we find
\[
\widetilde{\mathbb{Q}_0}
=
\frac{1}{2}
\left(
\frac{|f_{2}(p)|}{|f_{1}(p)|^2}
\right)^2
\]
and
\begin{equation}
\widetilde{\mathbb{Q}_1}
=
\frac{1}{|f_1(p)|^2}
\left(
\frac{|f_{2}(p)|^4}{|f_{1}(p)|^4}
\left(
\frac{11}{32}
-\frac{5}{12}
\mathrm{Re}
\left[
\frac{f_1(p)f_3(p)}{(f_2(p))^2}
\right]
\right)
+
\frac{1}{8}
\frac{|f_{3}(p)|^2}{|f_{1}(p)|^2}
\right).
\label{eq:leading_order_correction}
\end{equation}
There are two circumstances under which
$\widetilde{\mathbb{Q}}=\widetilde{\mathbb{Q}_0}=\widetilde{Q}$. The
first, which is the trivial case, is when the lattice spacing $|\xi|$
of the original lattice in the z-plane vanishes. The second case is
complex inversion $f(z)=1/z$ (or more generally a bilinear
transformation), for which we find
$\widetilde{\mathbb{Q}_1}=\widetilde{\mathbb{Q}_2}=\widetilde{\mathbb{Q}_3}...=0$.
The vanishing of all the higher order corrections means that all the
curves emanating from the vertex at $f(p)$ are circular arcs. It
follows that, for a 2D foam or a quasi 2D foam, Laplace's law is
satisfied by each of the arcs.

Let us now turn our attention to the mapping $f(z)=z^{1/2}$ in this
case we do not expect Laplace's law to hold for each of the arcs. We
have already computed the lowest order term in
Eq.($\!\!$~\ref{eq:w_mean_square_cur_cont}) and found that
\[
\widetilde{\mathbb{Q}_0}=Q=\frac{1}{2R^2}
\]
(see Eq.($\!\!$~\ref{eq:mean_square_analytical_1})). Upon substituting
$f(z)=z^{1/2}$ into Eq.($\!\!$~\ref{eq:leading_order_correction}) we
find
\[
\widetilde{\mathbb{Q}_1}=\frac{7}{128}\frac{1}{r^3}=\frac{7}{128}\frac{1}{R^6}.
\]
where we have transformed from the coordinates in the z-plane
$(r,\lambda)$ to the w-plane coordinates $(R,\Lambda)$ by using the
relationship $\sqrt{r}=R$. Assuming that the original lattice has a
constant lattice spacing $|\xi|$ we can say that a 2D foam, or a quasi
2D foam, is capable of producing a good approximation of a conformal
lattice when
\[
\widetilde{\mathbb{Q}_1}|\xi|^2=\frac{7}{128}\frac{|\xi|^2}{R^6}
\]
is small. We see that this happens at regions a large distance form
central (critical) point of the conformal lattice, i.e when
$R>>(7/128)^{1/6}|\xi|^{1/3}$.

{\bf Acknowledgements} DW acknowledges support from SFI, ESA and CNRS (visiting position at
the group of Prof M Adler, Univ. Paris-Est). AM is grateful to Trinity
College Dublin and Institute of Mathematical and Physical
Sciences, Aberystwyth University, for financial assistance for short visits including the 2007 winter meeting of the British Society of Rheology. AM and DW would like to thank M. A. Moore,  S. Cox and W. Drenckhan for advice and useful discussions.


\end{document}